\pdfoutput=1
\documentclass[twocolumn]{revtex4}
\usepackage{graphicx,epsfig}
\usepackage{amsmath}
\usepackage{amsfonts}
\usepackage{diagbox}
\usepackage{fancyhdr}
\usepackage{istgame}
\usepackage[pscoord]{eso-pic}

\usepackage{rotating}

\begin{document}



\pagestyle{fancy}
\lhead{\bf Network-Iterated Prisoner's Dilemma}
\rhead{Martín Soto Quintanilla}
\lfoot{Bachelor's Thesis}
\rfoot{Barcelona, June 2022}

\title{Network-Iterated Prisoner's Dilemma}
\author{Author: Martín Soto Quintanilla}
\affiliation{Facultat de F\'{\i}sica, Universitat de Barcelona, Diagonal
645, 08028 Barcelona, Spain.} 
\author{Advisor: 
Marián Boguñá Espinal}

\begin{abstract}
{\bf Abstract:} We introduce the stochastic Network-Iterated Prisoner's Dilemma (NIPD) model, a network of players playing the Prisoner's Dilemma with their neighbours, each with a memory-one strategy which they constantly and locally update to improve their success. This process is non-deterministic, and mirrors societal interactions in many relevant aspects. We use it to assess the flexibility, noise tolerance and real-world adaptability of some well-known strategies. Furthermore, in the model a new strategy naturally emerges which proves way more successful than those. We also derive some theoretical parameters that gauge the success of a strategy in this context.
\end{abstract}

\maketitle


\section{Introduction}
\vspace{-0,4cm}
The Prisoner's Dilemma is a simple two player game which abstractly models a certain kind of competitive social interaction. It is a foundational example in the theoretical field of Game Theory, and has found many applications in Economics or Rational Decision Theory in helping explain natural phenomena or empirical data.

In its original formulation, both players A and B are conceptualized as robbers just caught in a joint theft. They are put in separate rooms (can't communicate) and each has to decide whether to remain silent (cooperate) or betray the other. Of course, betraying the other yields personal gain, and so the years each player will have to serve in prison look like follows depending on their decisions:

\vspace{-0,2cm}

\begin{table}[h!]
\begin{tabular}{|c|l|l|}
\hline
\diagbox[innerwidth = 1.5cm, innerleftsep=0.3cm,innerrightsep=0.3cm]{\textbf{A}}{\textbf{B}}                                                      & \multicolumn{1}{c|}{\begin{tabular}[c]{@{}c@{}}B\\ cooperates\end{tabular}} & \multicolumn{1}{c|}{\begin{tabular}[c]{@{}c@{}}B\\ betrays\end{tabular}} \\ \hline
\begin{tabular}[c]{@{}c@{}}A\\ cooperates\end{tabular} &                  \diagbox[innerwidth = 1.5cm, innerleftsep=0.3cm,innerrightsep=0.3cm]{1}{1}                                                           &        \diagbox[innerwidth = 1.5cm, innerleftsep=0.3cm,innerrightsep=0.3cm]{5}{0}                                                                     \\ \hline
\begin{tabular}[c]{@{}c@{}}A\\ betrays\end{tabular}    &                             \diagbox[innerwidth = 1.5cm, innerleftsep=0.3cm,innerrightsep=0.3cm]{0}{5}                                                   &              \diagbox[innerwidth = 1.5cm, innerleftsep=0.3cm,innerrightsep=0.3cm]{3}{3}                                                               \\ \hline
\end{tabular}
\end{table}

\vspace{-0,2cm}

Notice thus that, if they act in a purely selfish and rational fashion, both A and B are always better off betraying, no matter the move the other player makes. The situation is paradoxical and problematic for our usual understanding of rationality, since both players acting rationally inhibits each of them from obtaining a better individual outcome through mutual cooperation. This is where the relevance of the Dilemma resides: it's a simple conceptualization of many real world situations in which individual rationality doesn't produce the best global outcome. It's been fruitfully applied to many such situations \cite{pris}.

In all generality, the payoffs of the table can be understood as any positive values that the players want to minimize, such as most prominently economical losses. Of course, the concrete numerical values can vary, and the same paradoxical situation will arise as long as the order relation between them remains the same (that is, it is no coincidence that we chose $5 > 3 > 1 > 0$). We stick to these values through this work.

So the situation is clear when playing a single instance of this game: rational players will inevitably defect. But the situation changes when the game is iterated through different rounds. In this situation, closer to the recurring interactions of the real world, the prospective gains of future cooperation can sometimes make betrayal an irrational decision. Reference \cite{Li} presents the mathematics of this Infinitely Iterated Prisoner's Dilemma as played by two players.

We will study it through a different treatment: we build a network of players repeatedly playing the Prisoner's Dilemma with their neighbors and constantly updating their individual strategies striving for better results. This is a more general model closer to how societal interactions and belief update really work. As we will see, it is more general and realistic for randomness to affect some aspects of the simulation, and so we are dealing with a stochastic process.

\vspace{-0,4cm}

\section{The NIPD model}
\vspace{-0,4cm}
We apply to our situation some ideas from the stochastic modelling of disease or information spreading, as presented in \cite{Hof}.

We build a network of $N = 10^3$ nodes (players), each one connected to some others (its neighbors) by a randomly generated adjacency vector (connections are of course bi-directional). To simplify treatment, we consider a degree-regular network, in which every node is connected to $k=4$ others.

Each one of the nodes has at all times a strategy. In all generality, this strategy can take into account any one of the previous played rounds. As a simplification, we consider those strategies depending only upon the outcome of the previous round. As presented in \cite{Li}, these memory-one strategies can be expressed as a 4-dimensional vector, in which every one of the 4 parameters expresses the probability the player has of cooperating in the next round, given the previous round having a certain of the 4 possible game outcomes. That is, if $C$ is cooperation and $B$ betrayal, each player's strategy will have the form

\vspace{2mm}
\centerline{$(p_{CC}, p_{CB}, p_{BC}, p_{BB})$}
\vspace{2mm}

\noindent where $p_{CC}$ is the probability of the player cooperating if the previous round resulted in double cooperation, $p_{CB}$ the probability of the player cooperating if the previous round they cooperated and were betrayed, etc. (and so all 4 parameters will be real numbers between 0 and 1).

As an illustrative example, a player with strategy $(1, 1, 1, 1)$ will always cooperate, and one with strategy $(0.5, 0, 0, 0)$ will only cooperate with a 50\% probability when the previous round resulted in double cooperation.

Notice this makes the network and process drastically more complex than that of mono-viral spread. There, the state of every node (healthy or infected) could be expressed by a binary value. Now we need four continuous variables.

We initialize the network by giving an initial strategy to every node. Then we start randomly selecting which neighboring nodes play a round. We could use continuous time as in \cite{Hof}, and give each connection between two nodes the parameter $\lambda_{ij}$ of a Poisson process that randomly determines when a round is played. But then we would choose all of these parameters equal as a simplification for tractability, and this is actually equivalent to just randomly choosing each time one connection (all connections having the same probability of being chosen). So we more simply use discrete time steps.

Of course, when a round is played, both players will choose one of the 4 parameters of their strategy depending on the result of the last round played between them, and then flip an unfair coin with that probability to determine whether they cooperate. \footnote{In the first round of the process, we choose players to act as if the previous (non-existent) round resulted in double cooperation, thus showing initial optimism or innocence. Choosing this initial condition differently doesn't drastically affect the initial dynamics.}

Based on the outcomes of the rounds, each player gradually accumulates their relative payoff $R_i$, a value which each player tries to minimize and serves as a measure for how good their strategy is doing. This value resets to 0 whenever the node or any of its neighbours changes strategy, and from then on accumulates the nodes' payoff minus the average of the neighbour's payoffs. That is, if $P_i^{(n)}$ is the accumulated payoff of a node $i$ since round $n$, and $i \sim j$ means the two nodes are connected, then 

\vspace{-0,6cm}

$$R_i = P_i^{(n)} - \frac{1}{k}\sum_{j\sim i}P_j^{(n)}$$ 

\vspace{-0,3cm}

\noindent where $n$ is the last round in which $i$ or one of its neighbours changed strategy.

After each played round, if the $R_i$ of any of the two players surpasses a threshold $R_{max} = 15$ \footnote{We could also decide when a node changes strategy by using any probability distribution on $R_i$, instead of a sharp cutoff $R_{max}$. The results obtained by most reasonable probability distributions don't seem to be drastically different.}, $i$ changes strategy. They do so by choosing their best performing neighbor (the one with lowest $R_j$) and copying their strategy \footnote{We also tried taking the average of their neighbour's strategies, but this yielded too thermalized systems rapidly converging to the absorbent state in which all nodes have the same strategy (which is just the average of all initial strategies).}. Notice this has as a consequence that any strategy a node has at any point of the process was already some node's strategy at the start of the process. The whole process is iterated for a certain number of played rounds. \footnote{Of course, our nodes play a big but finite amount of rounds. Strictly speaking, if a rational agent knows the number of rounds played will be finite, we will again be in the situation where it will always betray. But in our model we haven't given our players any information about the finiteness of the game, so we might conceptualize this as the players not knowing whether the game will end at any time.}

\vspace{-0,2cm}

\section{Four strategies}

\vspace{-0,2cm}

Dealing with all possible strategies is mathematically more complicated, so we start with a simpler situation where nodes can only have one of a small set of representative strategies (that furthermore have a clear motivation). Consider the following strategies \cite{Kuh}:

\begin{table}[h!]
\label{tab:my-table}
\begin{tabular}{|c|c|c|}
\hline
\textbf{Name}     & \textbf{Description}                                                                                          & \textbf{Vector} \\ \hline
Cooperator (C)    & Always cooperates                                                                                             & (1, 1, 1, 1)    \\ \hline
Traitor (T)      & Always betrays                                                                                                & (0, 0, 0, 0)    \\ \hline
Tit for Tat (TFT) & \begin{tabular}[c]{@{}c@{}}Copies the opponent's\\ last move\end{tabular}                                     & (1, 0, 1, 0)    \\ \hline
Pavlov (P)        & \begin{tabular}[c]{@{}c@{}}Cooperates only when\\ last round both players\\ played the same move\end{tabular} & (1, 0, 0, 1)    \\ \hline
\end{tabular}
\vspace{-4mm}
\end{table}

TFT is generally regarded as a versatile strategy, since it can protect itself from regular betrayal, but can also benefit from cooperation in the long run when possible. P on the other hand would not seem that smart a strategy, since for instance cooperating after being betrayed is hardly beneficial.

\begin{figure}[h!]
\centering

\includegraphics[width=\columnwidth]{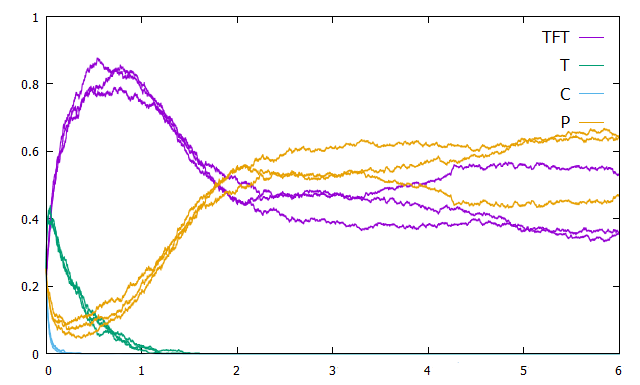}

\vspace{-0.1cm}

\footnotesize{$10^6$ rounds played}

\vspace{-0.2cm}

\caption{Percentile of the population carrying each of the four strategies, during three simulations of our model starting with equal (0.25) concentrations (with a same randomly chosen adjacency vector). T grows so rapidly to 0.4 that it's not appreciated due to scale.}
\label{fig:2}
\end{figure}

We first study the process beginning with equal amounts of the previous strategies distributed randomly across the population. This process is similar to a multi-viral scenario, where different viruses compete for population, but in our case the viral infection process is more complex than just accumulating viral load (it involves playing a game), and so different pairs of virus interact differently. The evolution of a realization of this process is presented in Fig.~\ref{fig:2}, which we discuss later.

Even in this case where strategies are non-probabilistic, the simulation is non-deterministic and the result does not only depend upon the initial conditions (the adjacency vector and the initial strategy distribution), but also on the specific realization of the simulation (that is, the random seed used to run the program). This is seen in all the successive figures, in which equal initial conditions yield different outcomes across different realizations. So every initial condition will have not one definite outcome, but an array of possible evolutions driven by different attractors, and thus a probability distribution of different final states.

From this we conclude that round order (the only random variable of the realization) can drastically affect the success of some strategies. This makes sense: for instance, a TFT with some Cs and some Ts as neighbors might by chance play some successive rounds against the Ts, with poor payoff which will lead to it changing strategy, or on the contrary against the Cs, from which it will benefit.

Exactly because of that, the only truly absorbent states in this situation are those in which all population carry the exact same strategy. This is because there's always a chance that a node updates strategy (although maybe very small), and if a neighbor has a different strategy this might entail an actual change. The existence of absorbent states implies the system to be non-ergodic (there are inescapable setups from which you can't access all phase space).

But there may be more states of stable equilibrium, even if they are not absorbent. Accurately assessing when these states appear would take way longer simulations, that our computing power doesn't allow. We could also quantitatively measure non-ergodicity by calculating the probability distribution of final absorbent or equilibrium states for a certain initial condition.

But we can still obtain information about the process from its starting dynamics and initial growths, as shown in Fig.~\ref{fig:2}, and there are some evident ways in which the initial conditions are relevant. In this setup, T grow rapidly at first, by taking advantage of the C and P. But T runs out of players to betray, and TFT then starts outperforming T (thanks to clusters of TFTs cooperating). In this newly non-betraying environment, the few P survivors start to thrive, and eventually reach a state of apparent oscillating equilibrium with higher concentration than TFT. This might really be a stable equilibrium in the long run, since it is conceivable that thanks to the topological distribution of connections two clusters of different strategies coexist. But it might also be destined, when enough rounds pass, to reach the absorbent state with all Ps or all TFTs (each with a certain probability of occurring in every simulation). That situation would be reminiscent of the Voter Model, where every realization can only end in one of two absorbent states, but the average of both these probabilities across realizations converges \cite{vote}.

We need a complementary theoretical parameter that objectively assesses how successful a strategy will be in a certain context. For that, we calculate the expected payoff of a strategy after each round (averaged over all its opponents). This will be a global variable approximating local behaviour. As an exemplification, consider the outcomes of iterated rounds between a TFT and a T at the start of the process:

\vspace{1mm}

\centerline{TFT vs T: CB  BB  BB  BB  ... [6]}

\noindent So, neglecting the exceptional first round, the expected payoff per round in the long run will be $E(\mbox{TFTvsT}) = E(\mbox{TvsTFT}) =3$ for both players. Applying the stochastic mean-field approximation, the probability of a node facing a certain strategy in a neighbor is just the overall concentration of that strategy in the whole population. Since at the beginning of this process these concentrations are all equal, the expected payoff per round of a TFT node will then be the following (and similarly we calculate the others):
\vspace{-0,2cm}$$E(\mbox{TFT}) = \frac{1}{4} (E(\mbox{TFTvsTFT}) + E(\mbox{TFTvsT}) + \ldots) = \frac{6}{4}$$\vspace{-0,8cm}$$E(\mbox{T}) = \frac{7.5}{4} \: \: \: \: \: \: \: \: E(\mbox{C}) = \frac{8}{4} \: \: \: \:\: \: \: \:E(\mbox{P}) = \frac{7}{4} $$ 

\vspace{-0,2cm}

This theoretically explains why TFT grows at the beginning, since its expected payoff is considerably lower (thus more successful) than all others. Still, behind this useful general assessment lie some topological and local details that cannot be captured by a single numerical value. How come, for instance, T grow at the expense of P, if $E(\mbox{T})$ is slightly bigger than $E(\mbox{P})$? This is because, even if overall T do slightly worse, any T surrounded by Cs or Ps will drastically benefit, and very rapidly transform its neighbors into other Ts.

We also notice that, in our model, the success of a strategy is heavily context dependent (a good sign that it approximates real situations). For instance, we see in Fig.~\ref{fig:2} that the dynamic changes when Cs and Ts are almost not present. Not only does the change in concentrations alter the previous expected payoffs, but furthermore the outcome of previous rounds will affect current performance. For instance, two previous Ts turned TFTs will continue forever betraying each other, failing to reconcile, to their disadvantage. On the other hand, PvsP will pass from any outcome to unending double cooperation. That's why in the aftermath, where the betraying T are no longer present, some P outperform some (non-cooperating) TFT.

This kind of beneficial noise tolerance or self-adjusting back to cooperation of P is not captured by the previous averaged payoff $E$, and so in some situations it might be more appropriate to consider a different average ($\bar{E}$) which also takes into account which of the possible outcomes of a game (CC, CB, BC, BB) might start a series of successive rounds between two players. As an exemplification, in our situation we can have, among others, the two sequences

\: \: \: \: \: \: \: \: \: \: \:  P vs P: CB  BB  CC  CC  ...

\: \: \: \: \: \: TFT vs TFT: BB  BB  BB  BB ...

\noindent (with longterm average payoff per round 1 and 3 respectively) and the more accurate averages will be
\vspace{-0,2cm}
$$\bar{E}(\mbox{PvsP}) = \frac{1}{4} (\bar{E}(\mbox{PvsP; CC}) + \bar{E}(\mbox{PvsP; CB}) + \ldots) = \frac{4}{4}$$
\vspace{-0,9cm}
$$\bar{E}(\mbox{TFT}) = \frac{16}{8} \: \: \: \: \: \: \: \: \bar{E}(\mbox{P}) = \frac{10}{8} $$

\vspace{-0,2cm}

\noindent which do show why P performs better. Notice this indicator will work best in chaotic situations where each strategy can come across different past outcomes (as in an ongoing process), and $E$ is more applicable when past outcomes are less chaotic (as in the start of a process).

\begin{figure}[h!]
\centering
\includegraphics[width=\columnwidth]{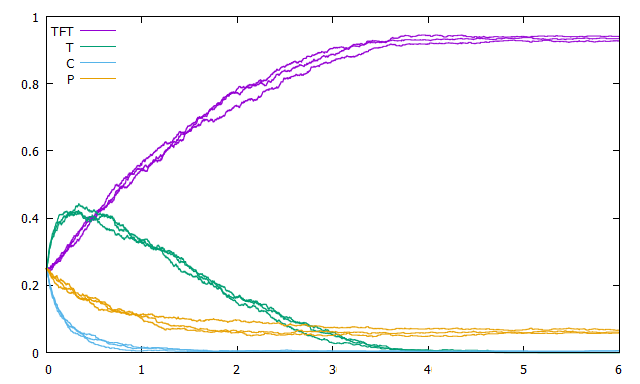}

\vspace{-0.1cm}

\footnotesize{$10^5$ rounds played}

\vspace{-0.2cm}

\caption{Percentile of the population carrying each of the four strategies, during three simulations of our model in which nodes act after a strategy change as if CC had just been played.}
\label{fig:3}
\end{figure}

TFT might perform better if they were somehow capable of making amends. One (maybe too artificial) way of ensuring this is making every node, when adapting a new strategy, act as if the last round was CC (thus every node can make amends). As we see in Fig.~\ref{fig:3}, this small memory reduction indeed ends the edge P had over TFT. It also yields way more clear hints of a stable equilibrium. The model with this simplification seems more tame.

Another (probably more natural) way is changing its strategy to be less drastically punitive. These are the $p$TFT of the next section.

\vspace{-0,8cm}

\section{One strategy against the world}
\vspace{-0,4cm}
We now deal with the richer complexity of probabilistic strategies, that is, vectors with any parameters.

One natural way of assessing the fitness of a strategy is seeing how well it does against any possible strategy. For that, we see how well a randomly scattered minority carrying a strategy does against a majority of individuals with randomly chosen strategies.

\begin{figure}[h!]
\centering
\includegraphics[width=\columnwidth]{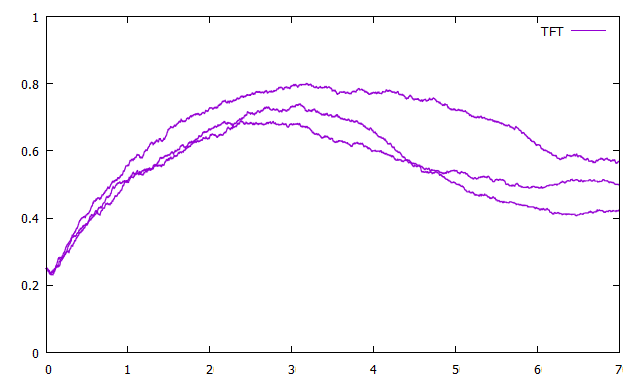}

\vspace{-0.1cm}

\footnotesize{$10^5$ rounds played}

\vspace{-0.2cm}

\caption{Percentile of the population carrying TFT, during three simulations of our model starting with $\frac{1}{4}$ of TFTs and $\frac{3}{4}$ random strategies (with a same randomly chosen adjacency vector and strategies).}
\label{fig:4}
\vspace{-0,3cm}
\end{figure}

We start with a quarter of the population initially carrying TFT. As seen in Fig.~\ref{fig:4}, TFT spreads fast (the posterior slight decay is probably due to other randomly appearing strategies more easily cooperating between them than TFT, as discussed earlier).

\begin{figure}[h!]
\centering
\includegraphics[width=\columnwidth]{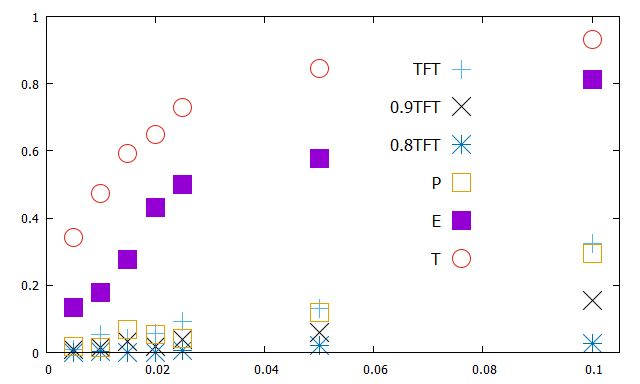}

\vspace{-0.1cm}

\footnotesize{initial concentrations ($\rho_i$)}

\vspace{-0.2cm}

\caption{Average over three simulations of the final concentrations ($\langle\rho_f\rangle$) of TFT, 0.9TFT, 0.8TFT, P, E (see next section) and T in a random environment (after $2\times 10^5$ rounds) for different initial concentrations of them ($\rho_i$). Error bars (derived from averaging) comparable to symbol size.}
\label{fig:6}
\end{figure}

We can study its pervasiveness (the fraction of nodes with a TFT strategy) after a certain number of rounds, depending on its initial concentration $\rho\in[0, 1]$. This is analogous to the study of viral endemicity \cite{Hof}, but now we can't lengthen the simulation enough to reach an equilibrium which confirms TFT remains endemic. But we can check the initial evolution (the concentration after a low number of rounds) for these values, as shown in Fig.~\ref{fig:6}, and theoretically reason that any $\rho$ for which the TFT don't die out quickly will present a stably TFT fraction of the population: the initial TFT cluster smaller than $\rho$ which always cooperates. So this is a highly local phenomenon. Basically, as long as there are enough TFTs so that by chance a cluster of them is formed, we will have both rapid growth of TFT, and stable permanence of at least some of them. We notice in Fig.~\ref{fig:6} that this happens even for very low concentrations, and so the concentration threshold above which we have endemicity is $\rho_t(\mbox{TFT}) \approx 0.005$. We also observe that a linear relation is apparent, although we would need many more realizations to average out the noise and provide moderately accurate parameters for the linear regression.

As mentioned last section, a less rigid form of TFT might be better equipped to deal with these probabilistic environments. So consider $p$TFT $= (p, p, 1-p, 1-p)$. Its performance (for two values of $p$) and those of other strategies are also presented in Fig.~\ref{fig:6}. We observe these variants of TFT actually perform worse. The context is very different from that of section III, and even if there TFT could have benefited from making amends, rigidity is more beneficial when facing random strategies. P also performs almost as well as TFT. Indeed its only relevant flaw was cooperating after CB, and that's not as big of a problem in a random environment where it won't usually encounter constant betrayers. T far outdoes all of them in initial growth. But when dealing with adequately long simulations, we do expect T to decay rapidly after a certain time (as in Fig.~\ref{fig:2}), since unlike P or TFT it can't form clusters of cooperation, and some random strategy probably will, to T's disadvantage.

Although not included due to lack of space, a more complex complementary theoretical assessment of the fitness of strategies (as before with $E$ and $\bar{E}$) can now be carried out by considering probabilistic game trees instead of sequences, and resulting equations.

\begin{figure}[h!]
\centering
\includegraphics[width=\columnwidth]{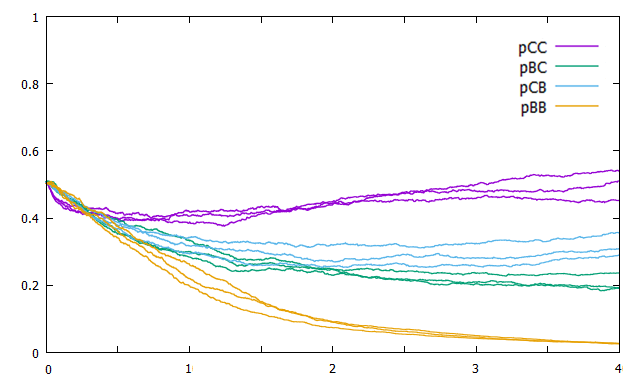}

\vspace{-0.1cm}

\footnotesize{$10^5$ rounds played}

\vspace{-0.2cm}

\caption{Evolution of the global average (for all nodes) of each of the four strategy parameters, during three simulations of our model starting with all random strategies.}
\label{fig:7}
\end{figure}

\vspace{-0,3cm}

\section{an emergent strategy}
\vspace{-0,4cm}
We might wonder what random strategies will prevail in a fully random environment. In that case we can't keep track of any one concrete strategy, so in Fig.~\ref{fig:7} we keep track of the global average of the strategy parameters of all nodes. This process rewards the strategies close to (taking the average approximately) E $=(0.5, 0.2, 0.35, 0)$. Might this be the strategy best fit to deal with a random environment? As we see in Fig.~\ref{fig:6}, it does indeed perform way better than the others, except for the unequaled starting growth of T. But we do know, as seen in Fig.~\ref{fig:7}, that this strategy performs better in the long run (unlike T), thanks to the benefit of some cooperation. So it does seem like an ideal all terrain strategy.

\vspace{-0,5cm}

\section{Conclusions}

\vspace{-0,3cm}

The NIPD model proves versatile and explanatory when dealing both with designed and random strategies. It provides a rich landscape, where many behaviors are heavily context dependent, mirroring some real societal dynamics. Thus it cannot always be analysed with complete mathematical accuracy.

The interaction between some of the strategies considered depend upon local phenomena and the topology of the network. Still, many macroscopic dynamics are easily noticed and intuitively explained. The two averages $E$ and $\bar{E}$ prove helpful tools (in different circumstances) for these macroscopic explanations.

The random strategies arena of section IV is a more pragmatic computational method to gauge a strategy's flexibility. We obtained interesting growth results, portraying the first moments of a strategy outbreak. Still, with more computational power, a more thorough study taking into account the pervasiveness of the strategy in far off stable equilibriums could be carried out.

The emergent strategy does overall great on this context. A more thorough sweep over the space of possible strategies (or averages taken as in Fig.~\ref{fig:7}) might provide slightly more well-adjusted parameters for an even more efficient strategy.

\vspace{-0,4cm}





\vspace{-0,3cm}
\begin{acknowledgments}
\vspace{-0,3cm}
I am immensely grateful to my supervisor Marián Boguñá for his help, patience, great ideas and great mood. It was a pleasure working with him.

Thank you also to all of my colleagues and friends of the double degree, who have made these five years the best of my life.
\end{acknowledgments}

\end{document}